# Room temperature ferromagnetism in intercalated Fe$_{3-x}$GeTe$_2$ van der Waals magnet

Hector Iturriaga[1], Luis M. Martinez[1], Thuc T. Mai[2], Mathias Augustin[3,3a], Angela R. Hight Walker[2], M. F. Sanad[4], Sreeprasad. T. Sreenivasan[4], Y. Liu[5$], Elton J. G. Santos[3,3b], C. Petrovic[6], Srinivasa R. Singamaneni[1*]

[1]Department of Physics, The University of Texas at El Paso, El Paso, Texas 79968, USA
[2]Quantum Metrology Division, Physical Measurement Laboratory, National Institute of Standards and Technology, Gaithersburg, Maryland 20899, USA
[3]Institute for Condensed Matter Physics and Complex Systems, School of Physics and Astronomy, The University of Edinburgh, Edinburgh EH9 3FD, UK
[3a]Donostia International Physics Centre (DIPC), Donostia-San Sebastian, 20018, Spain.
[3b]Higgs Centre for Theoretical Physics, The University of Edinburgh, EH9 3FD, UK
[4]Department of Chemistry, The University of Texas at El Paso, El Paso, Texas 79968, USA
[5]Los Alamos National Laboratory, MS K764, Los Alamos NM 87545, USA
[6]Condensed Matter Physics and Materials Science Department, Brookhaven National Laboratory, Upton, New York 11973, USA
$ Present address: Los Alamos National Laboratory, Los Alamos, New Mexico 87545, USA

**Abstract**

Among several well-known transition metal-based compounds, van der Waals (vdW) Fe$_{3-x}$GeTe$_2$ (FGT) magnet is a strong candidate for use in two-dimensional (2D) magnetic devices due to its strong perpendicular magnetic anisotropy, sizeable Curie temperature (T$_C$ ~ 154 K), and versatile magnetic character that is retained in the low-dimensional limit. While the T$_C$ remains far too low for practical applications, there has been a successful push toward improving it via external driving-forces such as pressure, irradiation, and doping. Here we present experimental evidence of a novel room-temperature (RT) ferromagnetic phase induced by the electrochemical intercalation of common tetrabutylammonium cations (TBA+) into FGT bulk crystals. We obtained Curie temperatures as high as 350 K with chemical and physical stability of the intercalated compound. The temperature-dependent Raman measurements in combination with vdW-corrected *ab initio* calculations suggest that charge transfer (electron doping) upon intercalation could lead to the observation of RT ferromagnetism. This work demonstrates that molecular intercalation is a viable route in realizing high temperature vdW magnets in an inexpensive and reliable manner.

*srao@utep.edu

**Introduction**

Since the discovery of long-range magnetic ordering in atomically thin CrI$_3$ and CrGeTe$_3$ magnets[1,2], there have been incredible efforts to unearth more compounds displaying 2D magnetism. [3–8] This discovery revitalized the discussion on the role of magnetic anisotropy in the Mermin-Wagner theorem[9], with recent results for finite systems demonstrated its limited applicability for real lab systems[10]. As a result, interest in the van der Waals (vdW) class of materials has skyrocketed[3-8], as there has been great success in uncovering ground-breaking findings in the fundamental physics of anisotropic low-dimensional magnetism.

Novel 2D materials are of particular interest because they create exciting opportunities for the realization of functional devices. It was found that bombarding graphene with protons induced ferromagnetism via single-atom defects, opening the door for opportunities in defect engineering of atomically thin vdW devices. [11-13] Since then, and after discovering a true 2D ferromagnet [1,2], there have been achievements in tuning magnetism, spin-phonon coupling, exchange interactions, superconductivity and novel charge and spin transport properties in vdW crystals and heterostructures stemming from proximity effects, high pressures, and mechanical defects. [3-8, 14–29] While developments in ultrathin vdW magnets are exciting, understanding of 2D limit characteristics stem from the bulk properties. Indeed, in $Mn_3Si_2Te_6$ and sister compound $Fe_{2.7}GeTe_2$ (FGT), we have previously reported strong changes in the magnetization, spin-phonon coupling and exchange interactions of vdW crystals after proton irradiation, photoexcitation, and pressure proving that external perturbations create an exciting environment for advantageously tuning the properties of vdW materials as the monolayer limit is approached. [14-16, 30-32]

In the bulk, FGT is a soft, layered ferromagnet within a vdW crystal structure. [33-36] It is part of a family of ternary transition metal tellurides, which are well known magnets for displaying 2D magnetism due to their strong structural stability after exfoliation. FGT displays a Curie temperature $T_C$ ~ 154 K, which, while being relatively high in comparison to the transition temperatures of similar compounds, is still far too low for any practical commercial application. Hence, efforts to increase the $T_C$ of FGT have included the intercalation of external molecules into its van der Waals gap. One report successfully observed a strong ferromagnetic ordering at high temperatures (>300 K) upon intercalation with sodium. [37] While exciting, that method created decomposition phases, effectively overpowering the intrinsic magnetism of FGT. Moreover, it did not provide results as transformative as that of tetrabutylammonium intercalation into $Cr_2Ge_2Te_6$. [38] In that report, there was an increase in $T_C$ of nearly 150 K, and a strengthening of the exchange mechanism. [38] Indeed, the literature proves that intercalation is a manageable and accessible method of altering the properties of layered materials. [17] Hence, we set out to intercalate FGT single crystals with tetrabutylammonium cations (TBA+) (Figure 1(a)) and quantify their effects on the magnetic properties.

In this work, we show that intercalated FGT exhibits room temperature ferromagnetism. By using temperature-dependent Raman measurements, in combination with vdW-corrected density functional theoretical calculations, we report $T_C$ values up to 350 K, which are robust against magnetic fields and temperature variations. The intercalation also modifies intrinsic properties of the FGT crystals including the spin-phonon coupling, the anisotropy field and uniaxial anisotropy coefficient. Our findings indicate that intercalated TBA+ molecules engineer the magnetic properties of FGT towards practical implementations.

**Results and Discussion**
Figure 1 (b) shows a comparison of the temperature-dependent magnetization of pristine and intercalated FGT. These measurements were conducted along the magnetic easy axis, where the field is applied perpendicular to the plane of the crystal, referred to as the out-of-plane direction (H//*c*).  For the zero-field cooled (ZFC) curves, the samples were cooled to 50 K in the absence of a magnetic field and allowed to thermally stabilize before applying a measuring field of 1 kOe and sweeping the temperature back up to 300 K. For the field-cooled (FC) curves, a cooling field of 1 kOe is applied as the sample is cooled and is maintained through the temperature sweep. In the case of ZFC measurements, the intercalated sample (denoted by black open circles) demonstrated a dramatic increase in magnetization when compared to the pristine crystal. Although the FC measurements between pristine and intercalated FGT (shown in solid green and black circles respectively) did not show much difference, the change in behaviour upon applying a cooling field is a clear indication that the magnetic response of intercalated FGT is altered upon intercalation.

The temperature-dependent magnetization data were used to estimate $T_C$ by taking the first derivative (see Supplementary Information). As shown in Figure 1, there is a negligible change in $T_C$ despite the changes observed in the magnetization. In fact, the values of the $T_C$ for both pristine and intercalated FGT agree well with the reported value of 153 K found in the literature. [33,35,36] In other words, the introduction of TBA+ did not affect the intrinsic $T_C$ of FGT, as is shown by the coinciding sharp drops in magnetization as the temperature is increased. Despite this, there is yet another striking difference in the temperature-dependent magnetization of intercalated FGT. In pristine FGT, after heating the sample through its $T_C$, there is a weaker magnetization as thermal fluctuations overpower the energy provided by the applied magnetic field. However, this is not true of intercalated FGT, which shows a significant magnetization even up to 300 K. This is a striking indication of long-range magnetic ordering sustained up to 300 K, which is atypical of iron deficient FGT such as the pristine samples we present. [32,35,36] Noting an initial drop in magnetization at $T_C \sim 153$ K, we attribute the high temperature behaviour of FGT to a secondary ferromagnetic phase induced by intercalation with TBA+ molecules.

This assertion is reinforced by isothermal magnetization data taken at low temperatures below and through $T_C \sim 153$ K up to 350 K. Shown in Figure 2, these 5-quadrant magnetization vs. magnetic field curves were taken at several temperature plots up to a maximum field of 30 kOe (3 T) in both polarities. As expected of soft-ferromagnetic FGT, there is small coercivity even for the intercalated sample. In addition, the relatively low saturation field indicates that the easy axis is indeed along the stacking direction (the crystallographic c-axis). Furthermore, this shows that the easy axis is stable upon intercalation. The most striking comparison between pristine and intercalated FGT is present at room temperature, where a clear ferromagnetic hysteresis is observed at room temperature only for the intercalated sample. In addition, these data present a "wasp-waisted" hysteresis which may be an indication of competition between two phases with firmly distinct coercivities. [39] Indeed, this effect is only seen after intercalation, since pristine FGT is clearly paramagnetic at 300 K as shown in Figure 2a (shown in black). Further evidence of a secondary phase is shown in Figure 2b, which presents a bifurcation of the 1st and 5th quadrant data from which a critical field of approximately 16 kOe (1.6 T) is extracted. This feature is not present in the data for the pristine FGT sample. Most importantly, in the temperature points near and above room temperature shown in Figure 2c, there is a clear hysteresis indicative of ferromagnetic ordering up to 350 K.

To gain deeper insights, we recorded the temperature evolution of the Raman spectra for the two phonon modes $E_{2g}^1$ and $A_{1g}^1$, which are related to the magnetic ordering of FGT. [34,40] First, in agreement with the literature, the wavenumber of these two modes shifts to lower values after intercalation (see Supplementary Information). [13,40] We propose that the effect is a result of not only weakened interlayer coupling stemming from increased separation distance of the FGT layers but also, and most importantly, by electron doping caused by intercalation. Indeed, Figure 3 includes the simulation results of the effect of TBA+ molecules on the charge rearrangement of FGT layers. A substantial amount of electrons from TBA+ are noticed into the Fe and Te atoms ranging from 0.16–0.13 electrons/layer and 0.51–0.72 electron/layer, respectively. The Ge atoms barely change their electronic charge with TBA+. In addition, we hypothesize that the changes in magnetic behaviour may correlate to alterations in the spin-phonon coupling of the intercalated sample. To investigate this possibility, Raman spectroscopy data in the parallel polarization configuration were collected at several temperatures and are shown in Figures 4a and 4b for pristine and intercalated FGT, respectively. The main phonon modes under investigation are the $E_{2g}^1$ and $A_{1g}^1$ modes with a Raman shift of approximately 125 cm$^{-1}$ and 142 cm$^{-1}$, respectively when exciting with 514.5 nm excitation. After modelling the spectra as a superposition of Lorentzian signals, the peak centres (in cm$^{-1}$) for each mode are plotted versus temperature on a logarithmic scale in Figure 4. There is a discontinuity in these values near T$_C$ ~ 153 K, which serves as an indication of spin-phonon coupling. [34,40] Hence, the points above the intrinsic transition temperature are fitted with a Boltzmann sigmoidal model to simulate the temperature-dependent behaviour of the phonon modes as though it were unperturbed by the ordering of the spins. This process is outlined in the previous report on the sister compound CrSiTe$_3$.[41] The clear deviation of our experimental data with the simulation at low temperatures is an additional evidence of spin-phonon coupling across both modes in pristine and intercalated FGT. This separation is quantified by taking the largest difference in the Raman shift ($\Delta\omega$) present at the lowest temperature (see Figure 4). This parameter is used to estimate the spin-phonon coupling parameter in the system.

Equation (1) is routinely used to extract the spin-phonon coupling parameter (λ').[41] However, equation (2) is the shorthand notation of equation (1). Finally, λ' is algebraically extracted using equation (3).

$$\omega^2 = \omega_0^2 + \lambda \langle \mathbf{S}_i \cdot \mathbf{S}_j \rangle \tag{1}$$
$$\omega \approx \omega_0 + \lambda' \langle \mathbf{S}_i \cdot \mathbf{S}_j \rangle \tag{2}$$
$$\lambda' = \frac{\Delta\omega}{\langle \mathbf{S}_i \cdot \mathbf{S}_j \rangle} \tag{3}$$

First, we take $\Delta\omega$ as the value extracted from the deviation of the sigmoidal model and the experimental data as the numerator. Then, we assume that all the magnetic contribution is a result of the Fe$^{3+}$ cation, [42] allowing us to set S$_i$ = S$_j$ = 3/2 and making the denominator equal to 9/4. From these values, the spin-phonon coupling parameters for both the $E_{2g}^1$ and $A_{1g}^1$ phonon modes are calculated for both pristine and intercalated FGT. The values are shown in Table 1. Our analysis shows that the spin-phonon coupling for the $E_{2g}^1$ mode decreases after intercalation. On the other hand, the opposite is true for the $A_{1g}^1$ mode, as λ' slightly increases after intercalation. Our analysis shows that there is a clear change in the spin-phonon coupling of FGT after intercalation. It is likely that this change is caused by alterations to the exchange interaction stemming from changes in the metal-ligand bond angles. [38].

To explain the presence of the room-temperature ferromagnetic phase, we look at the role of the intercalant molecule in the overall structure of FGT. To reiterate, we do not expect any substitution of TBA+ into the iron vacancies present in the sample. Therefore, we expect all TBA+ to reside in the van der Waals gap, creating an increased interlayer separation as reflected from the Raman data. With this, it is possible that the antiferromagnetic coupling between FGT layers is suppressed, leading to room-temperature ferromagnetism. [43–47] It was also found that electron doping has significant effects on the magnetic exchange within FGT. [46–48] In a previous report on TBA intercalation of Cr$_2$Ge$_2$Te$_6$, the electrochemical reaction responsible for the intercalation of TBA into the crystal, and an increase in the saturation magnetization show an evidence of electron doping in the sample. [38] Because we have intercalated our FGT with the same organic molecule, we assume that one full electron is added to FGT upon intercalation, as supported by our theoretical prediction (see below). It was shown that high levels of electron doping cause a drastic increase in T$_C$ by suppressing the AFM arising from the geometric frustration of inequivalent Fe sites. Suppression of the AFM coupling between FGT layers result in the most dramatic changes to T$_C$. [39] It is possible that a combination of both mechanical separation of the layers and doping-induced suppression of the AFM coupling is responsible for the creation of the room temperature magnetic ordering as the FM coupling is strengthened.

Another possible explanation lies in the electronic structure of intercalated FGT. $Cr_2Ge_2Te_6$ saw a dramatic increase in $T_C$ because of the occupation of conducting $d_{xz}/d_{yz}$ orbitals by the extra electrons introduced into the system upon intercalation. [38] This was reported due to a shift in exchange mechanism from a relatively weak superexchange across the ligand Te atom, to a robust direct exchange between Cr atoms. [38] The Stoner model readily describes the magnetic ordering of these systems, and the itinerant ferromagnetism of FGT is no exception. [42,33] Electron doping experiments on FGT have shown that the extra electrons begin to occupy the $d_{xz}/d_{yz}$ orbital subbands, and alter the electronic density of FGT, similar to intercalated $Cr_2Ge_2Te_6$. [41,42,49] As the Stoner model implies, changes in the density of states near the Fermi level would cause considerable changes to the ferromagnetic ordering of FGT. Therefore, we expect that intercalation-induced electron doping would explain room-temperature ferromagnetism by altering the electronic density of states of FGT in a way similar to intercalated $Cr_2Ge_2Te_6$, where there was a shift in the exchange mechanism to a strong Fe-Fe direct exchange facilitated by additional exchange channels provided by intercalants. [49]

Changes in the electronic density of states in $Cr_2Ge_2Te_6$ resulted in the complete switching of the magnetic easy axis. [29] While we did not detect such drastic changes in FGT, there is evidence that the anisotropy of the intercalated sample is decreased, further cementing the idea that the density of states of FGT near the Fermi level is affected in a similar manner to that of $Cr_2Ge_2Te_6$ after intercalation. An estimation of the anisotropy field and the uniaxial anisotropy constant of FGT reveal a decrease in both of these quantities after intercalation, which is detected as a decrease in the aforementioned anisotropy parameters. The results are shown in Figure 5. The mechanism at play may be the contribution of a planar orbital moment component allowing the spins to align along the in-plane direction more easily. [38] Indeed, intercalated FGT shows a smaller saturation field along the magnetic hard axis (H//ab), meaning that less energy is needed to overcome the anisotropy barrier and align the moments along their preferred direction. The anisotropy constant $K_u$ is related to the saturation magnetization $M_S$ and saturation field $H_{sat}$ along the in-plane direction by the Stoner-Wolfarth model:

$$\frac{2K_u}{M_S} = \mu_0 H_{sat} \tag{4}$$

where $\mu_0$ is the vacuum permeability (unity in cgs units). [42,33] The change in anisotropy is more readily correlated to a decrease in the saturation field, as the values of magnetization at high magnetic fields are comparable after intercalation. As a function of temperature, $K_u$ is consistently lower in the intercalated sample throughout the ferromagnetic phase. Theoretical calculations on electron doped FGT have also pointed to contributions from the $d_{xz}/d_{yz}$ orbitals as mediators of the anisotropy energy while having only a minimal effect on the magnetic moment per Fe atom. [42] In our intercalated FGT, we observe an appreciable decrease in the anisotropy, but no drastic changes in the high field magnetization. Therefore, we can expect that intercalated FGT sees a contribution from previously unoccupied states near the Fermi level and behaves similarly to $Cr_2Ge_2Te_6$ after intercalation. Thus, a shift in the exchange mechanism from a relatively weak superexchange to a stronger direct exchange may be what allows FGT to retain ferromagnetic ordering to such high temperatures.

**Conclusion**

To summarize, tetrabutylammonium cations were electrochemically intercalated into the vdW gap of $Fe_{2.7}GeTe_2$ bulk crystals. The effects were investigated via low temperature magnetometry and Raman spectroscopy measurements. Temperature-dependent magnetization data show a non-zero magnetization for intercalated FGT up to 350 K, confirmed by isothermal field-dependent magnetization data to be the result of an induced-ferromagnetic phase present far beyond the intrinsic $T_C \sim 154$ K of pristine FGT. Despite this, a transition is still observed at the intrinsic Curie temperature, and there are no changes in the saturation magnetization in either crystallographic orientation at low temperatures. However, above room temperature, the intercalated FGT shows a clear ferromagnetic hysteresis above at least 350 K.

The temperature-dependent Raman measurements coupled with our density functional theoretical calculations suggest that charge transfer could be the cause for the room temperature ferromagnetism upon intercalation. These findings lend strength to the idea of electrochemical intercalation as a viable method of realizing room-temperature magnetism for the purpose of incorporation into practical devices.

**Methods**

**Experimental measurements.** FGT crystals were grown through the self-flux technique, as outlined in previous reports, resulting in millimeter-sized single crystals.[34]
**Electrochemical Intercalation of FGT crystals with TBA$^+$.**
Initially, the FGT crystals were rinsed with ethanol many times. Then, single crystals were hard-pressed onto an indium plate and adjusted to work as positive electrode where 0.1 mm silver piece used as negative electrode. A saturated solution of tetrabutyl ammonium bromide powder (sigma aldrich, 99.0%) dissolved in DMF (sigma aldrich, 99.9%, was used for the electrochemical intercalation process. To perform homogenous intercalation, a potentiometric process was performed with a constant current of 2 µA. The whole intercalation process was carried out in an argon-medium using airtight cell. By applying the constant current, the negative electrode (Silver) will start to lose electrons and the FGT crystals will start to accept those electrons. The amount of intercalated material is then determined from the discharge curve and the specific criteria set by Wang et al. [38]

X-ray diffraction data (see Supplementary Information) were taken for the pristine and intercalated samples. This XRD data helps to confirm that the intercalant molecules indeed reside in the van der Waals gap of the crystal. This is of particular concern as this FGT sample is iron deficient, leaving a wide opportunity for molecules to readily substitute into any vacancies instead of intercalating into the van der Waals gap. In other words, we are confident that any changes to the magnetic behaviour of our samples intercalated with TBA+ molecules are not due to the destruction of the intrinsic crystal structure of FGT. As reported, if the amount of intercalant exceeds an optimal value, then the structure of the crystal would be completely disrupted. Due to the sharp peaks and correspondence to previous literature reports, we can assert that intercalation did not affect the structure or crystallinity of FGT. [33,35,50]

Low-temperature bulk magnetometry measurements for intercalated FGT were conducted using a Quantum Design Physical Property Measurement System with the AC Measurement System option. DC moment data were extracted in a temperature range extending from 5 K to 350 K and a maximum magnetic field of ± 30 kOe. Orientation dependent measurements were conducted for both the in plane (H//ab) and out of plane (H//c) direction of the magnetic field with respect to the crystallographic c-axis of the crystal.

Raman spectroscopy measurements were performed in the parallel polarization configuration on the bulk crystals. A fresh surface was exfoliated for the measurements. To ensure a reduction of noise in the signals, the samples were exfoliated. Measurements were conducted in the temperature range of 3.2 K to 295 K.

**vdW corrected ab initio simulations.** First-principles simulations were done using the Vienna Ab-initio Simulation Package (VASP) [51], with the PAW pseudo-potentials using the Perdew–Burke–Ernzerhof (PBE) exchange-correlation and an energy cutoff of 600 eV. We needed to use a 2x2x1 FGT supercell to intercalate the TBA. This led to a supercell containing 154 ions. We used the DFT-D3 corrections [52] to account for the van der Waals interactions between the FGT layers and TBA molecules. The relaxations were done using convergence criteria of $10^{-8}$ eV and a k-mesh of 1x1x1 at the Gamma point. The single-shot calculations were done using convergence criteria of $10^{-9}$ eV and a k-mesh of 1x1x1 at the Gamma point to compute the charge density. The single-shot calculations were done on the full system (TBA+FGT), and on two subsystems constructed by removing either the FGT or TBA from the full system so that we could compute the charge density difference via $\Delta \rho = \rho(FGT + TBA) - \rho(FGT) - \rho(TBA)$.

**Acknowledgements:** This material is based upon work supported by the National Science Foundation Graduate Research Fellowship Program under Grant No. 184874.1 Any opinions, findings, and conclusions or recommendations expressed in this material are those of the author(s) and do not necessarily reflect the views of the National Science Foundation. S.R.S. and H.I. acknowledge support from the NSF-DMR (Award No. 2105109). SRS acknowledges support from NSF-MRI (Award No. 2018067). E.J.G.S. acknowledges computational resources through CIRRUS Tier-2 HPC Service (ec131 Cirrus Project) at EPCC (http://www.cirrus.ac.uk) funded by the University of Edinburgh and EPSRC (EP/P020267/1); ARCHER UK National Supercomputing Service (http://www.archer.ac.uk) via Project d429. E.J.G.S. acknowledges the Spanish Ministry of Science's grant program "Europa-Excelencia" under grant number EUR2020-112238, the EPSRC Early Career Fellowship (EP/T021578/1), and the University of Edinburgh for funding support. Work at Brookhaven National Laboratory was supported by the U.S. Department of Energy, Office of Basic Energy Science, Division of Materials Science and Engineering, under Contract No. DE-SC0012704 (crystal growth).

**Figures:**

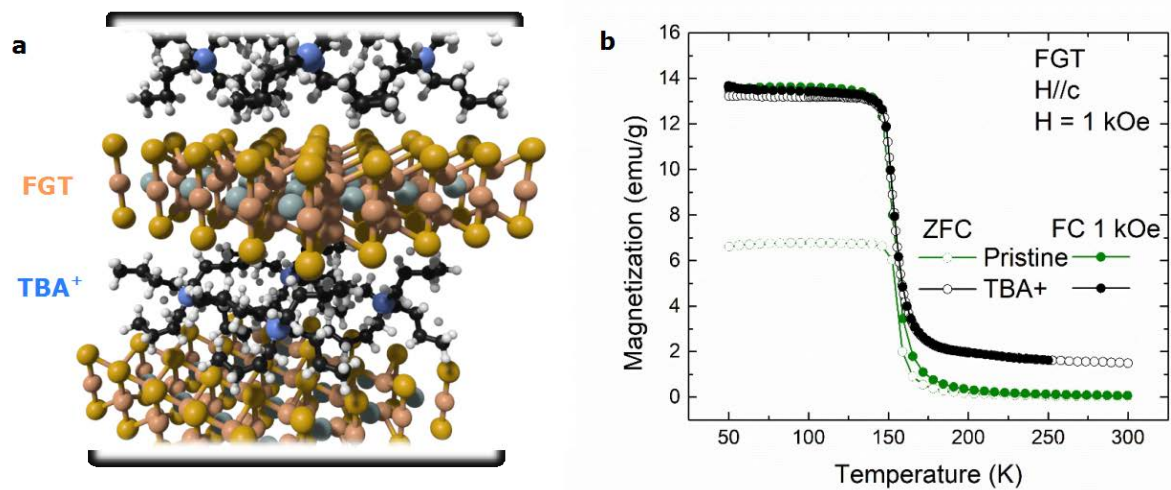

***Figure 1. a,*** *Atomic scheme of FGT and TBA+ molecules intercalation.* ***b,*** *Magnetization vs. Temperature curves for pristine (green) and intercalated (black) FGT. Zero-field cooled (open symbols) and field-cooled (solid symbols) curves are also shown. The data are taken in the out of plane direction.*

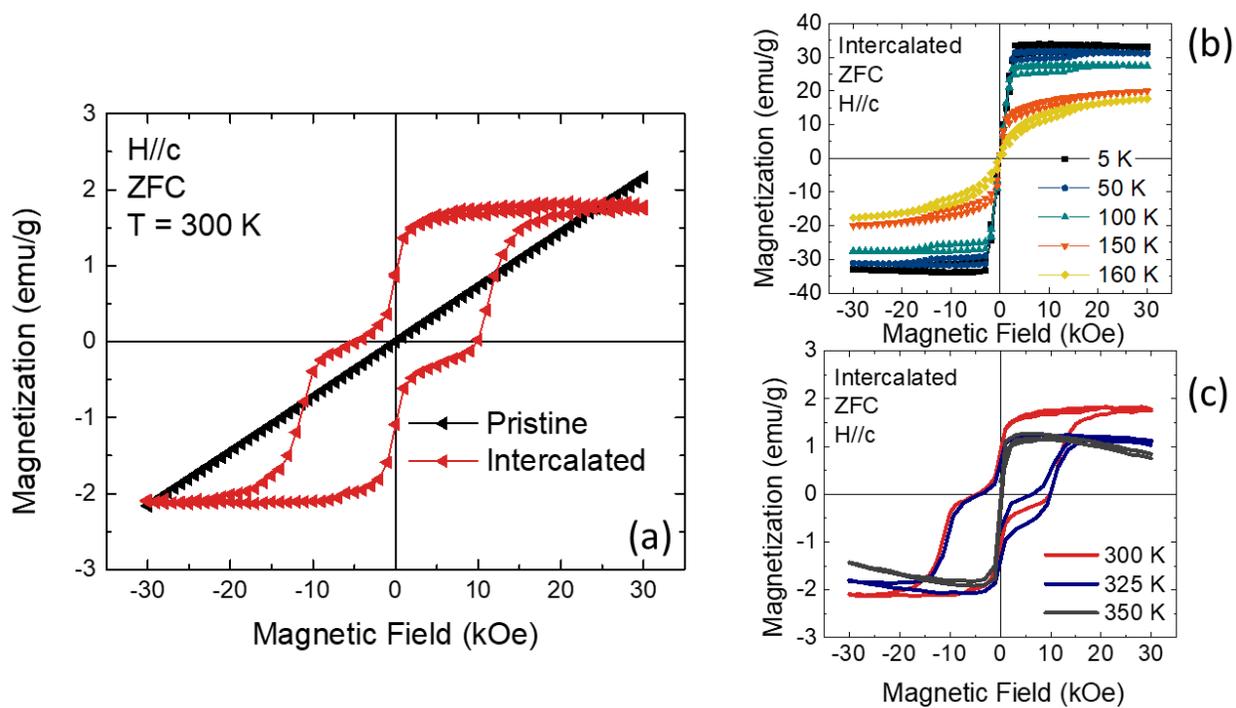

*Figure 2.* Isothermal magnetization shown at several temperatures. Figure 2a shows a comparison between fully paramagnetic pristine (black) and the room-temperature ferromagnetic behavior of intercalated (red) FGT. Figure 2b and 2c display the ferromagnetic ordering of intercalated FGT up to 350 K.

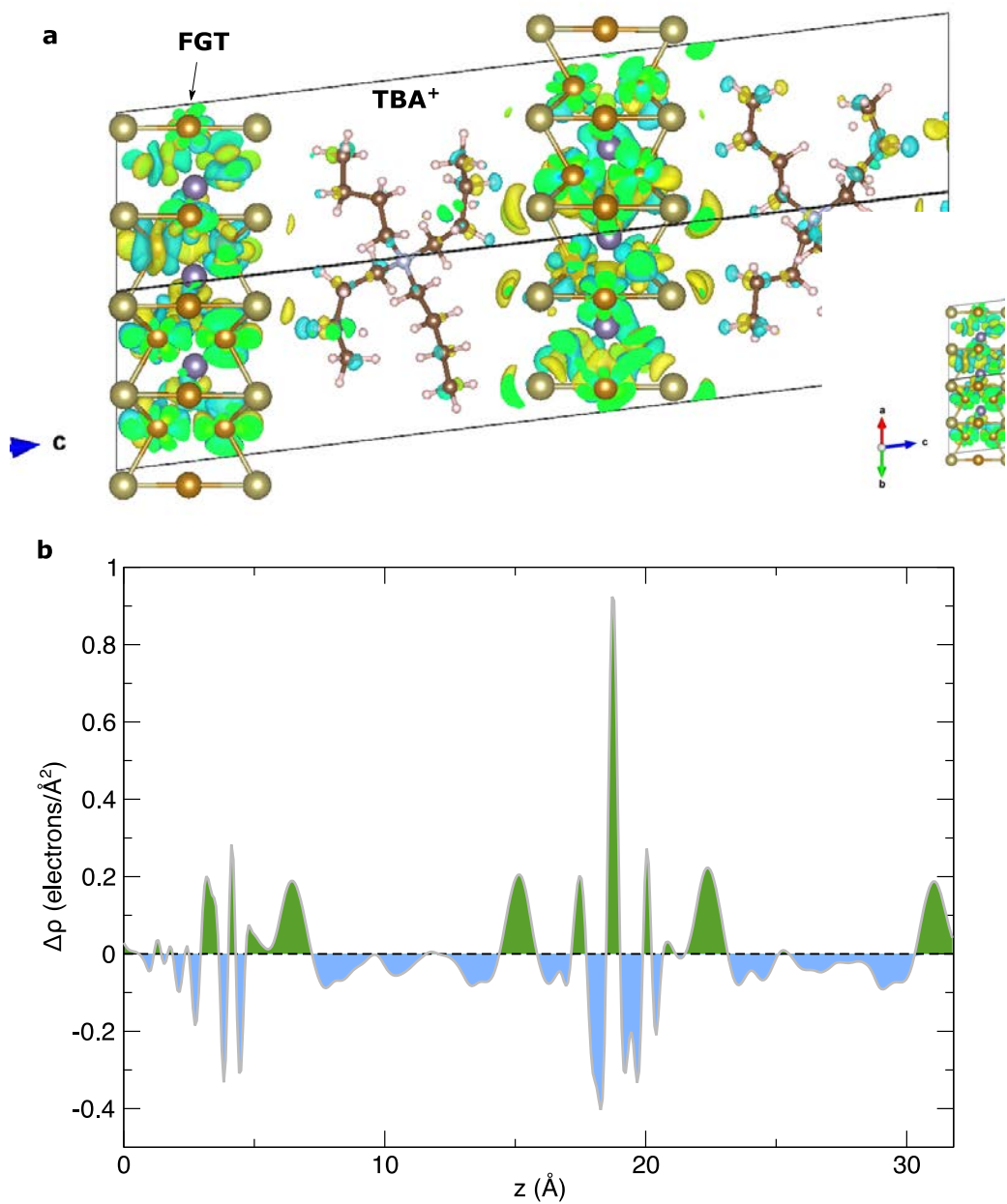

*Figure 3. a-b*, Charge density difference ($\Delta\rho = \rho(FGT + TBA) - \rho(FGT) - \rho(TBA)$, where $\rho(FGT + TBA)$ is the total charge density of FGT plus TBA, $\rho(FGT)$ is the charge density of the isolated FGT system, and $\rho(TBA)$ is the charge density of the TBA molecules) through iso-surface and planar average along the z direction, respectively. In **a**, green and blue iso-surfaces correspond to positive and negative charge, respectively

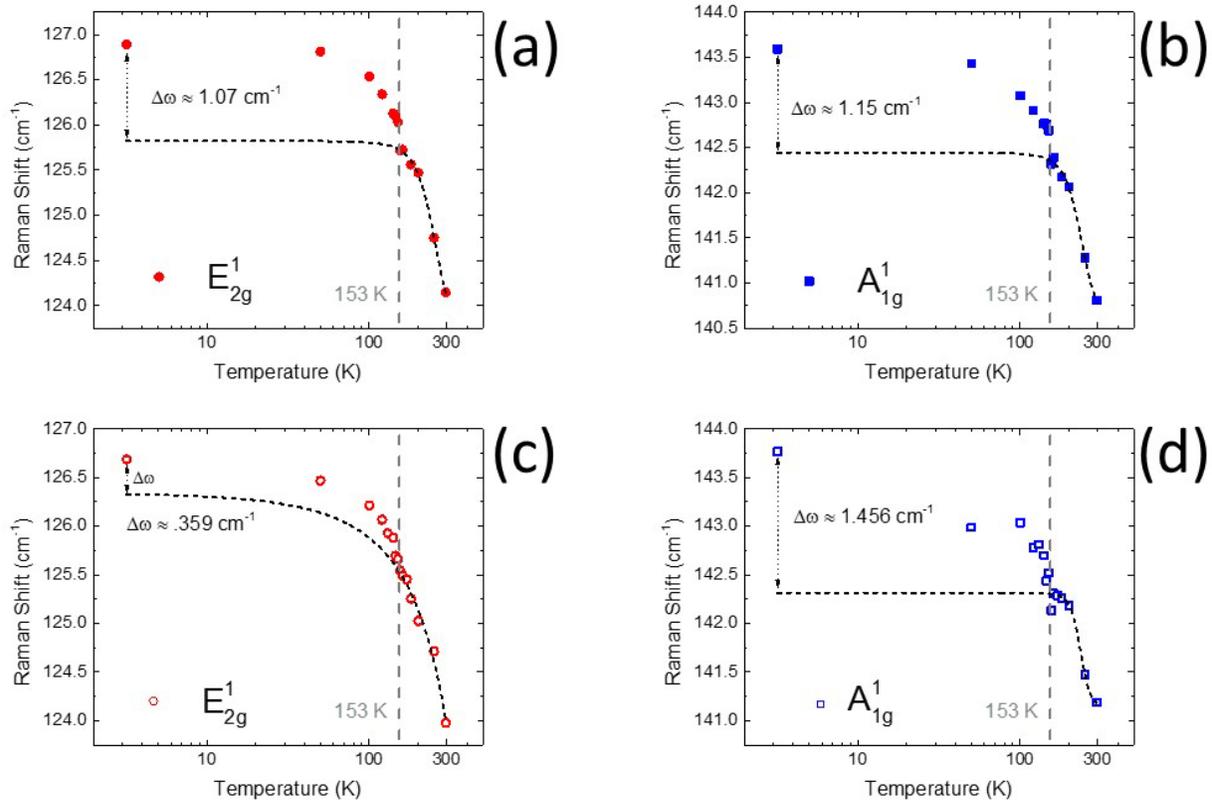

*Figure 4.* Peak centers for both phonon modes as a function of temperature. The temperature evolution of these modes for pristine (closed symbols, figures a,b) and intercalated (open symbols, figures c,d) FGT is modeled using a Boltzmann sigmoidal model (dashed line). The deviation between this simulated unperturbed mode and the experimental data is readily apparent.

|  |  | Δω (cm$^{-1}$) | λ' (cm$^{-1}$) |
|---|---|---|---|
| $E_{2g}^1$ | Pristine | 1.07 | .476 |
|  | Intercalated | .359 | .160 |
| $A_{1g}^1$ | Pristine | 1.15 | .511 |
|  | Intercalated | 1.456 | .647 |

*Table 1 Comparison of the spin-phonon coupling parameters for both pristine and intercalated FGT.*

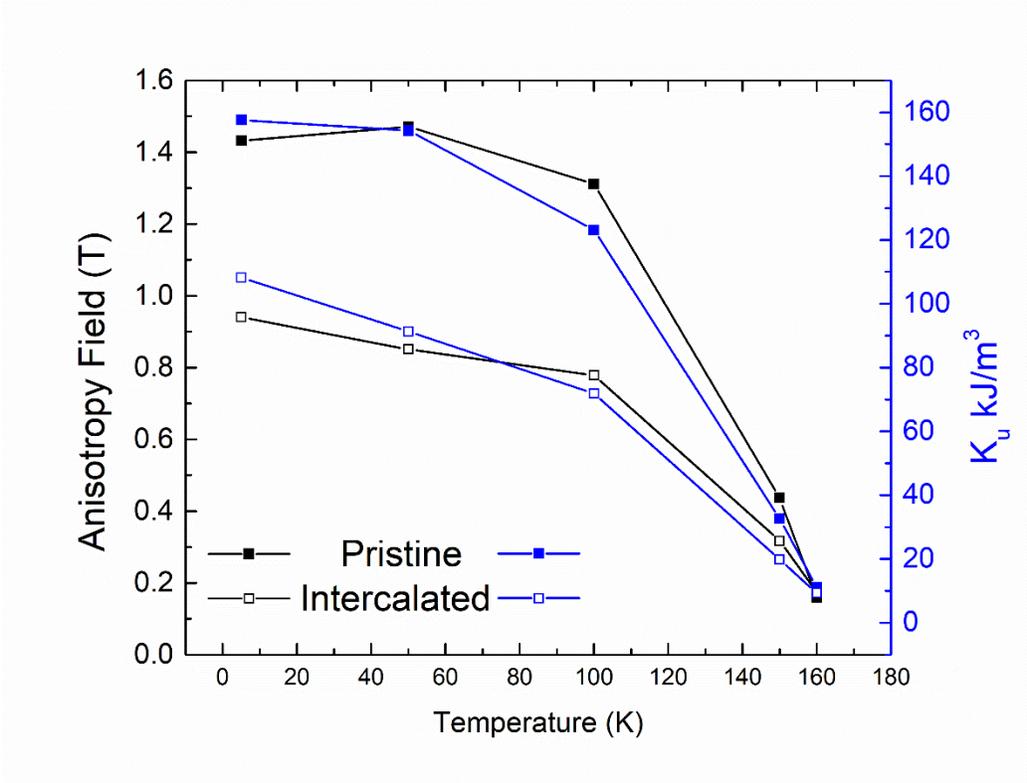

*Figure 5. The anisotropy field (black, left y-axis) and the uniaxial anisotropy coefficient (blue, right y-axis) are shown for both pristine (solid symbols) and intercalated (open symbols) in the low temperature region. It is shown that the anisotropy parameters of FGT decrease noticeably after intercalation.*

# Supplementary Information

**Room temperature ferromagnetism in organic molecule-intercalated $Fe_{3-x}GeTe_2$**

Hector Iturriaga[1], Luis M. Martinez[1], Thuc T. Mai[2], Mathias Augustin[3,3a], Angela R. Hight Walker[2], M. F. Sanad[4], Sreeprasad. T. Sreenivasan[4], Y. Liu[5$], Elton J. G. Santos[3,3b], C. Petrovic[6], Srinivasa R. Singamaneni[1*]

[1]Department of Physics, The University of Texas at El Paso, El Paso, Texas 79968, USA
[2]Quantum Metrology Division, Physical Measurement Laboratory, National Institute of Standards and Technology, Gaithersburg, Maryland 20899, USA
[3]Institute for Condensed Matter Physics and Complex Systems, School of Physics and Astronomy, The University of Edinburgh, Edinburgh EH9 3FD, UK
[3a]Donostia International Physics Centre (DIPC), Donostia-San Sebastian, 20018, Spain.
[3b]Higgs Centre for Theoretical Physics, The University of Edinburgh, EH9 3FD, UK
[4]Department of Chemistry, The University of Texas at El Paso, El Paso, Texas 79968, USA
[5]Los Alamos National Laboratory, MS K764, Los Alamos NM 87545, USA
[6]Condensed Matter Physics and Materials Science Department, Brookhaven National Laboratory, Upton, New York 11973, USA
$ Present address: Los Alamos National Laboratory, Los Alamos, New Mexico 87545, USA
Laboratory, Upton, New York 11973, USA
*srao@utep.edu

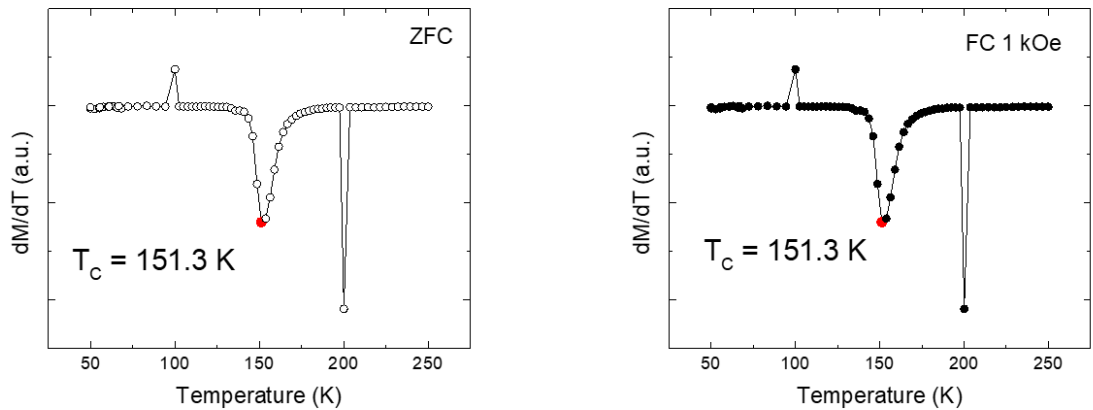

**S1.** First derivative of the zero field-cooled (open symbols) and field-cooled (closed symbols) magnetization with respect to the temperature for the intercalated FGT. The minimum observed at 151.3 K approximates the onset of the transition from ferromagnetism to paramagnetism in the sample. The extrema at 100 and 200 K are artifacts from the magnetometer and should be ignored.

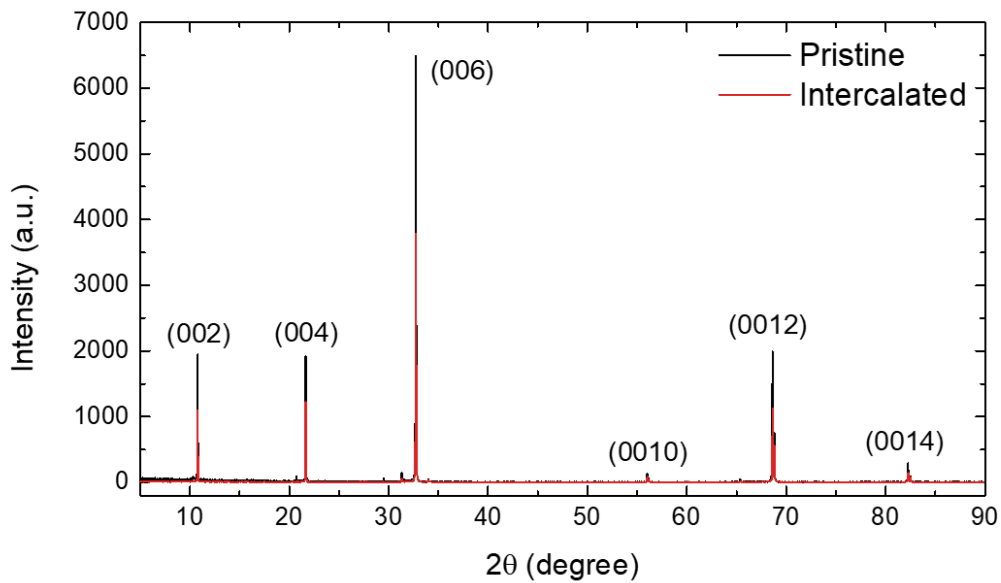

**S2.** PXRD spectra for pristine (black) and intercalated (red) FGT.

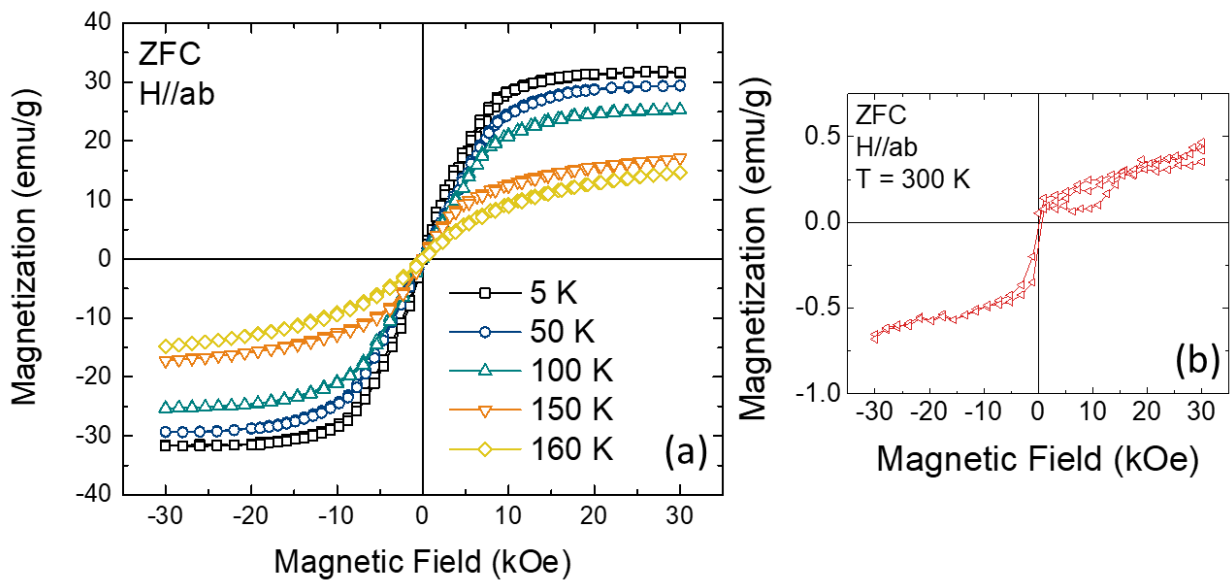

**S3.** In plane magnetization vs. magnetic field for intercalated FGT up to 300 K (b). Ferromagnetism persists even along the hard axis at high temperatures.

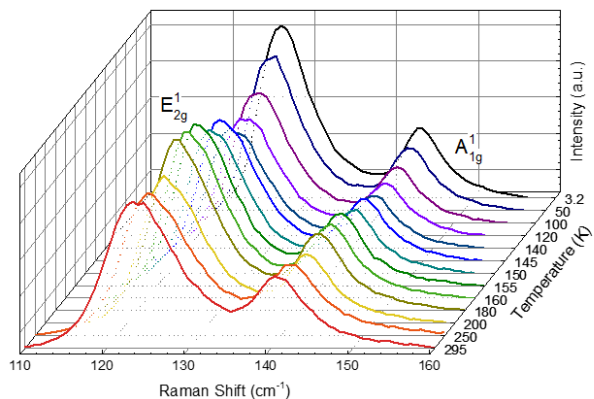
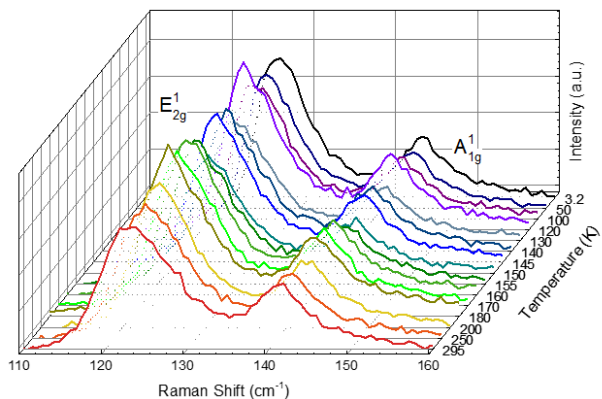

**S4.** Raman spectra for both pristine (a) and intercalated FGT (b) at several temperatures ranging from 3.2 K to 295 K. The measurements were done in the parallel configuration with a wavelength of 515 nm.